\documentclass[
superscriptaddress,
nofootinbib,
 amsmath,amssymb,
 aps,
pra,
twocolumn]{revtex4-2}

\usepackage[utf8]{inputenc}
\usepackage{graphicx}
\usepackage{hyperref}
\hypersetup{colorlinks = true, linkcolor = [rgb]{0.19411,0.51882,0.667058}, urlcolor = [rgb]{0.125490,0.29542,0.1647058}, citecolor = [rgb]{0.75882,0.37411,0.14117}, breaklinks=true}
\usepackage{amsmath}
\usepackage{amsfonts}
\usepackage{amssymb}
\usepackage{dsfont}
\usepackage{braket}
\usepackage{comment}
\usepackage[format=plain]{caption}
\usepackage{subcaption}
\newcommand{\added}[1]{\textcolor{black}{#1}} 
\newcommand{\addedd}[1]{\textcolor{black}{#1}} 
\newcommand*{\I}{\mathrm{i}\mkern1mu}

\begin{document}

\title{Interferometric detection of continuous-variable entanglement using two states}

\author{Elena Callus}\email{elena.callus@uni-jena.de}
\affiliation{Institute of Condensed Matter Theory and Optics, Friedrich-Schiller-Universit\"{a}t Jena, Max-Wien-Platz 1, 07743 Jena, Germany}
\author{Martin G\"{a}rttner}\email{martin.gaerttner@uni-jena.de}
\affiliation{Institute of Condensed Matter Theory and Optics, Friedrich-Schiller-Universit\"{a}t Jena, Max-Wien-Platz 1, 07743 Jena, Germany}
\author{Tobias Haas}\email{tobias.haas@ulb.be}
\affiliation{Centre for Quantum Information and Communication, \'{E}cole polytechnique de Bruxelles,
CP 165, Universit\'{e} libre de Bruxelles, 1050 Brussels, Belgium}

\begin{abstract}
The efficient witnessing and certification of entanglement is necessitated by its ubiquitous use in various aspects of quantum technologies. In the case of continuous-variable bipartite systems, the Shchukin--Vogel hierarchy gives necessary conditions for separability in terms of moments of the mode operators. In this work, we derive mode-operator-based witnesses for continuous-variable bipartite entanglement relying on the interference of two states. Specifically, we show how one can access higher moments of the mode operators, crucial for detecting entanglement of non-Gaussian states, using a single beamsplitter with variable phase and photon-number-resolving detectors. We demonstrate that the use of an entangled state paired with a suitable reference state is sufficient to detect entanglement in, e.g., two-mode squeezed vacuum, NOON states, and mixed entangled cat states. We also take into account experimental noise, including photon loss and detection inefficiency, as well as finite measurement statistics.
\end{abstract}

\date{\today}

\maketitle

\section{Introduction}

Entanglement is a genuine quantum phenomenon, whereby states share correlations that are stronger than what is possible in the classical realm \cite{Bell1964}. The rise in quantum technologies, which rely on entanglement as a critical operational resource \cite{Ekert1991,Pittman1995,Wootters1998,Jozsa2003}, has sustained a strong interest in characterizing and efficiently certifying entanglement \cite{Plenio2007,Horodecki2009,Ghne2009}. A wide range of different entanglement criteria have been proposed. These are presented as criteria that all separable states must adhere to and the violation of which implies the presence of entanglement. A common theme underpinning many of the aforementioned criteria is the violation of the positivity of the partially transposed state, or positive partial transpose (PPT) criterion \cite{Peres1996,Horodecki1996}.

The advance of continuous-variable quantum systems as a viable platform for quantum technologies \cite{Braunstein2005,Kok2007,Weedbrook2012,Serafini2017, Slussarenko2019} has necessitated the understanding of entanglement in the context of infinite-dimensional Hilbert spaces. This has led to the development of, e.g., moments-based \cite{Duan2000,Simon2000,Mancini2002,Giovannetti2003,Agarwal2005,Hertz2016} and entropic \cite{Walborn2009,Walborn2011,Saboia2011,Tasca2013,Schneeloch2018,Haas2021b,Haas2022a} entanglement criteria, as well as entanglement witnesses based on continuous majorization theory \cite{,Haas2023a,Haas2023b}. The Shchukin--Vogel hierarchy \cite{Shchukin2005}, which is what we will be considering in this work, was proposed as a way to extend the characterization of single-mode nonclassicality using an infinite hierarchy of conditions based on observable moments of the bosonic mode operators \cite{Shchukin2005a,Shchukin2005b} to bipartite entanglement. It is a reformulation of the PPT criterion, with the hierarchy providing necessary and sufficient conditions for this to hold.

Although various formulations of entanglement criteria have been proposed, their practical efficiency is not obvious from an experimental point of view. In the case of small-dimensional Hilbert spaces, reconstruction of the state by means of tomography is the optimal method for certification \cite{James2001,Cramer2010}. However, this has been proven to be extremely inefficient for continuous-variable systems, except for Gaussian states due to their simple structure \cite{Mele2024}. Instead, the use of multiple copies of the state has been proposed as a more feasible way of flagging entanglement. The use of state replicas was first introduced as a way of estimating functionals of the density matrix \cite{Ekert2002,Alves2003,Brun2004}. This has then been further developed to constrain phase-space uncertainty \cite{Hertz2019}, optical non-classicality \cite{Arnhem2022,Griffet2023b} and entanglement \cite{Griffet2023b,Deside2025}, and has been experimentally demonstrated in, for example, cold-atom setups \cite{Daley2012,Islam2015,Kaufman2016}. However, the requirement of multiple identical copies poses experimental challenges. Entanglement generation in optical states is typically probabilistic, such as in the case of linear optical networks \cite{Knill2001,Zhang2018,Pankovich2024} and spontaneous parametric down-conversion \cite{Couteau2018}, or limited by difficulties in perfect realization \cite{Lee2019,Vezvaee2022}. Therefore, scaling the number of concurrently produced identical states translates to very low success probabilities. 

In this work, we address this obstacle by proposing a measurement scheme to detect entanglement using only two interfering states: the entangled state and an auxiliary state, which does not need to be a replica of the former. By measuring the photon-number distribution after interference, we estimate moments of the mode operators which demonstrate violation of separability criteria belonging to the Shchukin--Vogel hierarchy. We demonstrate that our two-state scheme is sufficient for the certification of entanglement in Gaussian, two-mode Schr\"{o}dinger cat, Hermite--Gaussian, and NOON states, even when using a (separable) product coherent state as the auxiliary.

The paper is organized as follows. In Section~\ref{sec:shchukinvogel}, we provide the reader with a brief introduction to the Shchukin--Vogel hierarchy. Next, in Section~\ref{sec:measurementscheme}, we describe the two-state measurement scheme and show how this can be used to experimentally demonstrate the violation of a certain class of separability criteria. This is followed by an extension of the criteria to the general case of non-identical state copies. In Section~\ref{sec:examplestates} we benchmark the performance of our scheme, assuming a coherent state as the reference, for a range of classes of entangled states, and in Section~\ref{sec:experimentalconsiderations} we discuss the impact of experimental noise and finite measurement statistics. Finally, we provide a summary of our work and discuss potential future directions in Section~\ref{sec:conclusions}.

\section{Separability criteria}\label{sec:separabilitycriteria}

In this section, we will start by introducing the Shchukin--Vogel hierarchy. Following this, we will present the scheme for experimentally accessing the relevant moments using an arbitrary auxiliary state.

\subsection{Shchukin--Vogel hierarchy}\label{sec:shchukinvogel}

We consider bipartite continuous-variable quantum systems that can be described by means of bosonic field operators $\boldsymbol{a}$ and $\boldsymbol{b}$, which satisfy the bosonic commutation relations $[\boldsymbol{a},\boldsymbol{a}^\dagger]=[\boldsymbol{b},\boldsymbol{b}^\dagger]=\mathds{1}$. The Peres--Horodecki, or positive partial transpose (PPT), criterion states that all separable states have a non-negative partial transpose, i.e., $\boldsymbol{\rho}^\text{PT}\geq 0$ \cite{Peres1996,Horodecki1996}. Hence, the violation of this inequality is a sufficient (but not necessary) condition for entanglement. 

In an attempt to characterize PPT in continuous-variable systems in terms of observable quantities, the Shchukin--Vogel hierarchy was derived \cite{Shchukin2005}, with further developments in \cite{Miranowicz2006,Miranowicz2009}. It consists of a hierarchy of conditions expressed in terms of observable moments of the mode operators, which collectively is necessary and sufficient for PPT. Specifically, for any normal-ordered operator $\boldsymbol{f}$ of the form $\boldsymbol{f}=\sum_{n,m,k,l}c_{nmkl}\boldsymbol{a}^{\dagger n}\boldsymbol{a}^m\boldsymbol{b}^{\dagger k}\boldsymbol{b}^l$, where $c_{nmkl}\in\mathbb{C}$, a separable state $\boldsymbol{\rho}$ must satisfy the condition
\begin{equation}\label{eq:criterion}\begin{split}
  &\braket{\boldsymbol{f}^\dagger\boldsymbol{f}}^\text{PT}\\  &\,=\sum_{\substack{n,m,k,l\\p,q,r,s}}c^*_{pqrs}c_{nmkl}\braket{\boldsymbol{a}^{\dagger q}\boldsymbol{a}^{p}\boldsymbol{a}^{\dagger n}\boldsymbol{a}^{m}\boldsymbol{b}^{\dagger s}\boldsymbol{b}^{r}\boldsymbol{b}^{\dagger k}\boldsymbol{b}^{l}}^\text{PT}\geq 0  ,
\end{split}\end{equation}
where $\braket{\cdot}^\text{PT}$ denotes the expectation value with respect to the partially-transposed state. By Sylvester's criterion, Eq.~\eqref{eq:criterion} holds if and only if all its principal minors are non-negative. Note that in the case of continuous-variable systems, the partial transpose operation corresponds to $\boldsymbol{b}\rightarrow\boldsymbol{b}^\dagger$ \cite{Simon2000}, which implies that $\braket{\boldsymbol{a}^{\dagger q}\boldsymbol{a}^{p}\boldsymbol{a}^{\dagger n}\boldsymbol{a}^{m}\boldsymbol{b}^{\dagger s}\boldsymbol{b}^{r}\boldsymbol{b}^{\dagger k}\boldsymbol{b}^{l}}^\text{PT}=\braket{\boldsymbol{a}^{\dagger q}\boldsymbol{a}^{p}\boldsymbol{a}^{\dagger n}\boldsymbol{a}^{m}\boldsymbol{b}^{\dagger l}\boldsymbol{b}^{k}\boldsymbol{b}^{\dagger r}\boldsymbol{b}^{s}}$.

\subsection{Measurement scheme}\label{sec:measurementscheme}

\begin{figure}
    \centering
    \includegraphics[width=0.85\linewidth]{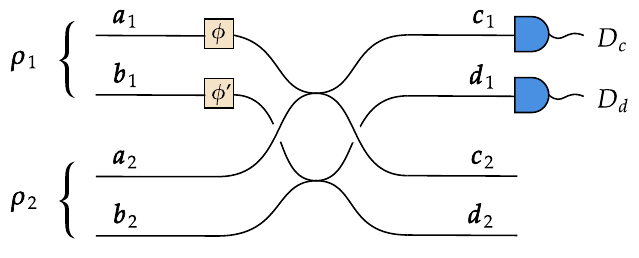}
    \caption{A schematic of the measurement setup consisting of two balanced beamsplitters with variable phase-shifts $\phi$ and $\phi'$, and photon-number-resolving detectors $D_c$ and $D_d$ at the upper output arms. The input state is given by $\boldsymbol{\rho}=\boldsymbol{\rho}_1\otimes\boldsymbol{\rho}_2$, and we interfere mode $\boldsymbol{a}_1$ with $\boldsymbol{a}_2$ and mode $\boldsymbol{b}_1$ with $\boldsymbol{b}_2$, respectively.}
    \label{fig:setup}
\end{figure}

Next, we discuss the measurement scheme that can be implemented in order to detect potential violation of separability conditions derived from the Shchukin--Vogel hierarchy. We consider the setup depicted schematically in Fig.~\ref{fig:setup}, comprised of two balanced beamsplitters, with variable phase-shifts, and photon-number-resolving (PNR) detectors. The input state is a separable product state $\boldsymbol{\rho}=\boldsymbol{\rho}_1\otimes\boldsymbol{\rho}_2$, where $\boldsymbol{\rho}_i$ describes a bipartite state made up of modes $\boldsymbol{a}_i$ and $\boldsymbol{b}_i$, for $i=1,2$. We consider the general case where $\boldsymbol{\rho}_1\neq\boldsymbol{\rho}_2$. The transformation resulting from the interference between the modes $\boldsymbol{a}_1$ and $\boldsymbol{a}_2$, and modes $\boldsymbol{b}_1$ and $\boldsymbol{b}_2$, is given by
\begin{equation}\label{eq:modetransformation}
    \begin{pmatrix}
        \boldsymbol{a}_1\\
        \boldsymbol{b}_1\\
        \boldsymbol{a}_2\\
        \boldsymbol{b}_2\\
    \end{pmatrix} \longrightarrow \begin{pmatrix}
        \boldsymbol{c}_1\\
        \boldsymbol{d}_1\\
        \boldsymbol{c}_2\\
        \boldsymbol{d}_2\\
    \end{pmatrix} = \frac{1}{\sqrt{2}}\begin{pmatrix}
        \boldsymbol{a}_1 e^{\I\phi}+\boldsymbol{a}_2\\
        \boldsymbol{b}_1 e^{\I\phi'}+\boldsymbol{b}_2\\
        \boldsymbol{a}_1 e^{\I\phi}-\boldsymbol{a}_2\\
        \boldsymbol{b}_1 e^{\I\phi'}-\boldsymbol{b}_2\\
    \end{pmatrix},
\end{equation}
and the PNR detectors measure the output at modes $\boldsymbol{c}_1$ and $\boldsymbol{d}_1$. 

This allows us to measure correlation functions of the form $\braket{(\boldsymbol{c}_1^\dagger\boldsymbol{c}_1)^{m'}(\boldsymbol{d}_1^\dagger\boldsymbol{d}_1)^{n'}}$, which can be expressed in terms of the input modes:
\begin{equation}\begin{split}\label{eq:nocorrelation}
    &\braket{(\boldsymbol{c}_1^\dagger\boldsymbol{c}_1)^{m'}(\boldsymbol{d}_1^\dagger\boldsymbol{d}_1)^{n'}}\\
   &=\frac{1}{2^{m'+n'}}\langle(\boldsymbol{a}_1^\dagger\boldsymbol{a}_1+\boldsymbol{a}^\dagger_2\boldsymbol{a}_2+\boldsymbol{a}^\dagger_1\boldsymbol{a}_2e^{-\I\phi}+\boldsymbol{a}^\dagger_2\boldsymbol{a}_1e^{\I\phi})^{m'}\\
    &\qquad \qquad \quad \times(\boldsymbol{b}_1^\dagger\boldsymbol{b}_1+\boldsymbol{b}^\dagger_2\boldsymbol{b}_2+\boldsymbol{b}^\dagger_1\boldsymbol{b}_2e^{-\I\phi'}+\boldsymbol{b}^\dagger_2\boldsymbol{b}_1e^{\I\phi'})^{n'}\rangle.
\end{split}\end{equation}
The photon-number correlator defined here is therefore a Fourier series where all the coefficients are linear combinations of moments of order $m'+n'$. The coefficients of the largest frequency terms (i.e., \added{coefficients of $e^{\pm\I (m'\phi+n'\phi')}$ and $e^{\pm\I (m'\phi-n'\phi')}$}) \added{can then be extracted by performing a Fourier analysis of the photon-number correlation as a function of the phase shifts $\phi$ and $\phi'$ in the interferometric setup.} Therefore, we are able to access the following set of moments:
\begin{equation}\label{eq:set}
    \left\lbrace \braket{\boldsymbol{a}_i^{\dagger m'}\boldsymbol{b}_i^{\dagger n'}}_i\langle\boldsymbol{a}_j^{ m'}\boldsymbol{b}_j^{ n'}\rangle_j,\braket{\boldsymbol{a}_i^{\dagger m'}\boldsymbol{b}_i^{ n'}}_i\braket{\boldsymbol{a}_j^{ m'}\boldsymbol{b}_j^{\dagger n'}}_j \right\rbrace_{\substack{i,j=1,2 \\ \added{i\neq j}}}\;,
\end{equation}
where $\braket{\cdot}_i$ denotes the expectation value with respect to the state $\boldsymbol{\rho}_i$. Furthermore, we have used the fact that the two initial states, $\boldsymbol{\rho}_1$ and $\boldsymbol{\rho}_2$, are uncorrelated, and hence $\braket{\boldsymbol{f}(\boldsymbol{a}_1,\boldsymbol{b}_1)\,\boldsymbol{g}(\boldsymbol{a}_2,\boldsymbol{b}_2)}=\braket{\boldsymbol{f}(\boldsymbol{a}_1,\boldsymbol{b}_1)}_{1}\braket{\boldsymbol{g}(\boldsymbol{a}_2,\boldsymbol{b}_2)}_{2}$.

Next, we move on to the entanglement witnesses. We consider separability criteria derived from second-order minors in the Shchukin--Vogel hierarchy of the form
\begin{equation}\begin{split}\label{eq:som}
    0\leq d_{mnpq}=&\braket{\boldsymbol{a}^{\dagger m}\boldsymbol{a}^{m}\boldsymbol{b}^{\dagger n}\boldsymbol{b}^{n}}\braket{\boldsymbol{a}^{\dagger p}\boldsymbol{a}^{p}\boldsymbol{b}^{\dagger q}\boldsymbol{b}^{q}}\\
    &\,-|\braket{\boldsymbol{a}^{\dagger m}\boldsymbol{a}^p\boldsymbol{b}^{\dagger q}\boldsymbol{b}^{n}}|^2,
\end{split}\end{equation}
where only one element in the sets $\lbrace m,p\rbrace$ and $\lbrace n, q\rbrace$ is non-zero. In the case of interference of two replica states, i.e., when $\boldsymbol{\rho}_1=\boldsymbol{\rho}_2$, \added{we have that $\boldsymbol{a}_1=\boldsymbol{a}_2$ and we can, hence, drop the indices $i,j$ such that the second summand in Eq.~\eqref{eq:som} corresponds to one of the terms in Eq.~\eqref{eq:set}.}

We note that the first summand is a product of moments, each of which has an equal number of creation and annihilation operators for the two modes. This can be reordered into a linear combination of moments of the photon number operators:
\begin{equation}\label{eq:firstterm}
    \braket{\boldsymbol{a}^{\dagger m}\boldsymbol{a}^{m}\boldsymbol{b}^{\dagger n}\boldsymbol{b}^{n}}=\sum_{k,l} C_{kl}^{mn}\braket{(\boldsymbol{a}^{\dagger }\boldsymbol{a})^{k}(\boldsymbol{b}^{\dagger }\boldsymbol{b})^{l}},
\end{equation}
where the coefficients $C_{kl}^{mn}$ may be expressed in terms of Stirling numbers of the second kind.\footnote{The photon-number operator raised to the $n$th power in normally-ordered form is given by $(\boldsymbol{a}^\dagger\boldsymbol{a})^n=\sum_{k}S(n,k)\boldsymbol{a}^{\dagger k}\boldsymbol{a}^k$, where $S(n,k)=\sum_{i=0}^k\frac{(-1)^{k-i}i^n}{{k-i)!i!}}$ are Stirling numbers of the second kind \cite{Katriel1992}.} The right-hand side of Eq.~\eqref{eq:firstterm} is composed of moments of the photon-number distribution of the original state $\boldsymbol{\rho}_i$, with $i=1,2$.  Hence, the minor $d_{mnpq}$ can be estimated by measuring photon-number correlations of the state with and without interference.\footnote{We remark that full photon-number resolution is not strictly necessary for sufficiently low-ordered moments. In such cases, photon-detection capabilities that allow for an accurate estimation of the relevant moments of the photon-number distribution are sufficient.}

To overcome the assumption of copies, we introduce a new separability condition $d'_{mnpq}$, which is based on the second-order minor, $d_{mnpq}$, in Eq.~\eqref{eq:som} and makes use of the measurable moments. Without loss of generality, suppose $n=p=0$ and let $\boldsymbol{\rho}_\epsilon=\boldsymbol{\rho}_1-\boldsymbol{\rho}_2$. Note that $\boldsymbol{\rho}_\epsilon$ is traceless and Hermitian by definition. Then, we obtain the separability condition
\begin{equation}\label{eq:modmin}\begin{split}
    d&'_{mnpq}\\
    &=\frac{1}{2} \sum_{\substack{i\neq j}}\left(\braket{\boldsymbol{a}_i^{\dagger m}\boldsymbol{a}_i^{m}}_i\braket{\boldsymbol{b}_i^{\dagger q}\boldsymbol{b}_i^{q}}_i-\braket{\boldsymbol{a}_i^{\dagger m}\boldsymbol{b}_i^{\dagger q}}_i\braket{\boldsymbol{a}_j^{ m}\boldsymbol{b}_j^{ q}}_j\right)\\
    &=\frac{1}{2}\left(d_{mnpq}^{(1)}+d_{mnpq}^{(2)}+\braket{\boldsymbol{a}^{\dagger m}\boldsymbol{b}^{\dagger q}}_\epsilon\braket{\boldsymbol{a}^{ m}\boldsymbol{b}^{ q}}_\epsilon\right) \geq 0,
\end{split}\end{equation}
where the last term is real and non-negative by Hermiticity of $\boldsymbol{\rho}_\epsilon$ and $d_{mnpq}^{(i)}$ is the value of the minor with respect to state $\boldsymbol{\rho}_i$. \added{The condition that exactly one element in the sets $\lbrace m,p \rbrace$ and $\lbrace n,q \rbrace$ is zero applies here as well.} This inequality is satisfied when \emph{both} $\boldsymbol{\rho}_1$ and $\boldsymbol{\rho}_2$ are separable, and hence its violation is a sufficient condition to flag entanglement in at least one of the two states. 

It is clear to see that when using two replicas of a state, i.e. when $\boldsymbol{\rho}_1=\boldsymbol{\rho}_2$, we recover the original condition $d_{mnpq}\geq 0$. On the other hand, one may choose to pair the entangled state with some suitable reference state. Throughout this work, we choose the reference to be the product coherent state $\ket{\gamma,\delta}$, with complex-valued coherent state amplitudes $\gamma$ and $\delta$. This state saturates all inequalities given by Eq.~\eqref{eq:som}, i.e. $d^{\ket{\gamma,\delta}}_{mnpq}=0$ for all $\gamma,\delta$. Furthermore, there exists an optimal coherent state, which depends on the entangled state and the criterion of choice, that minimizes $d'_{mnpq}$ and recovers the original minor. This occurs when the moment $\braket{\boldsymbol{a}^{\dagger m}\boldsymbol{a}^p\boldsymbol{b}^{\dagger q}\boldsymbol{b}^n}$ with respect to the entangled and the product coherent state are equal. For example, setting $n=p=0$ as was done in Eq.~\eqref{eq:modmin}, we would then obtain
\begin{equation}
   \braket{\boldsymbol{a}^{\dagger m}\boldsymbol{b}^{\dagger q}}_\epsilon=\braket{\boldsymbol{a}^{\dagger m}\boldsymbol{b}^{\dagger q}}_1-\braket{\boldsymbol{a}^{\dagger m}\boldsymbol{b}^{\dagger q}}_{2}=0,
\end{equation}
and therefore $d'_{mnpq}=\frac{1}{2}d_{mnpq}$. \addedd{We outline to experimental procedure to estimate the minor $d'_{mnpq}$ in Eq.~\eqref{eq:modmin} in the following step-by-step procedure:
\begin{enumerate}
    \item Estimate the moments $\braket{\boldsymbol{a}^{\dagger m}\boldsymbol{a}^{m}}$ and $\braket{\boldsymbol{b}^{\dagger q}\boldsymbol{b}^{q}}$ with respect to states $\rho_1$ and $\rho_2$ from the photon count distribution observed without any state-interference. These terms correspond to the first summand of Eq.\eqref{eq:modmin}.
    \item Using the interferometric setup, estimate the photon-number correlators $\braket{(\boldsymbol{c}_1^{\dagger}\boldsymbol{c}_1)^{m}(\boldsymbol{d}_1^{\dagger}\boldsymbol{d}_1)^{q}}$ as a function of the phase shifts $\phi,\phi'$. The moment $\braket{\boldsymbol{a}_i^{\dagger m}\boldsymbol{b}_i^{\dagger q}}_i\braket{\boldsymbol{a}_j^{m}\boldsymbol{b}_j^{q}}_j$, where $i,j=1,2$ can be extracted by Fourier analysis, as outlined in Eq.~\eqref{eq:nocorrelation}. This constitutes the second summand of Eq.\eqref{eq:modmin}.
    \item Calculate the value of $d'_{mnpq}$ from the previous steps, whereby a negative value certifies the presence of entanglement.
\end{enumerate}}

We remark that all separability conditions considered here are phase-insensitive, i.e., invariant under local rotations. This is evident upon noting that the summands of the minor have the same number of creation and annihilation operators. Therefore, any rotation transformation, which is described by $\boldsymbol{a}\rightarrow \boldsymbol{a}e^{\I\theta}$ and $\boldsymbol{a}^\dagger \rightarrow \boldsymbol{a}^\dagger e^{-\I\theta}$ and similarly for mode $\boldsymbol{b}$, leaves the minor unchanged. Physically, this means that entanglement can be detected regardless of the state's rotational orientation in phase space.

Although the criterion in Eq.~\eqref{eq:modmin} is weaker than Eq.~\eqref{eq:som}, we will show that it is sufficient to detect several classes of entangled states. Moreover, this allows us to relax the requirement of multiple replicas of the entangled state to just a product of the entangled state and some suitable auxiliary, or reference, state which can be chosen to be separable. Finally, the scheme has the advantage that the physical setup is the same for all orders of moments, requiring only modifications in the statistical analysis post-measurement and adjustments in the number of shots to account for the sampling complexity of the underlying distribution.

\section{Example states}\label{sec:examplestates}

Next, we will benchmark the performance of the two-state measurement scheme by considering several relevant families of entangled states. For each class of entangled states, we will derive an appropriate entanglement witness $d'_{mnpq}$ and show that entanglement can be flagged when employing a suitable reference state. Throughout this section, we will consider implementation of the scheme with the product coherent state $\ket{\gamma,\delta}$, with coherent state amplitudes $\gamma$ and $\delta$, as the reference state.

\subsection{Two-mode squeezed vacuum}

We start with the Gaussian state of two-mode squeezed vacuum. This is described in the Fock basis by
\begin{equation}
    \ket{\Psi_\text{TMSV}}=\sqrt{1-\lambda^2}\sum_{n=0}^{\infty}\lambda^n\ket{n,n},
\end{equation}
where $\lambda\in(-1,1)$ and the state is entangled for all $\lambda\neq 0$. We choose as a suitable separability criterion the following:
\begin{equation}\label{eq:tmsv}\begin{split}
    d'_{1001}&=\frac{1}{2}\left(d_{1001}+\braket{\boldsymbol{a}^\dagger\boldsymbol{b}^\dagger}_\epsilon\braket{\boldsymbol{a}\boldsymbol{b}}_\epsilon\right)\\
    &=\frac{1}{2}\left(\frac{-\lambda^2}{1-\lambda^2}+\left|\frac{\lambda}{1-\lambda^2}-\gamma\delta\right|^2\right)\geq 0,
\end{split}\end{equation}
\added{where
\begin{equation}
    d_{1001}=\braket{\boldsymbol{a}^\dagger\boldsymbol{a}}\braket{\boldsymbol{b}^\dagger\boldsymbol{b}}-\braket{\boldsymbol{a}^\dagger\boldsymbol{b}^\dagger}\braket{\boldsymbol{a}\boldsymbol{b}}=-\frac{\lambda^2}{1-\lambda^2}.
\end{equation}
}Note that $d_{1001}$ is strictly negative for all $\lambda\neq 0$. The witness $d'_{1001}$ is not invariant under local displacements, even in the case of state replicas, which are represented by the operator transformations $\boldsymbol{a}\rightarrow \boldsymbol{a}+\alpha$ and $\boldsymbol{b}\rightarrow \boldsymbol{b}+\beta$ with $\alpha$ and $\beta$ denoting the amount of displacement of the respective modes. Indeed, the authors in Ref.~\cite{Griffet2023b} show that a third-order minor and three state replicas are required to ensure that the witness is invariant under displacements. Nonetheless, the separability criterion is violated for a suitable auxiliary state when the following constraint is satisfied:
\begin{equation}
    \lambda(|\alpha|^2+|\beta|^2-1)<(\alpha\beta+\alpha^*\beta^*).
\end{equation}

\begin{figure}
    \centering
    \includegraphics[width=1.0\linewidth]{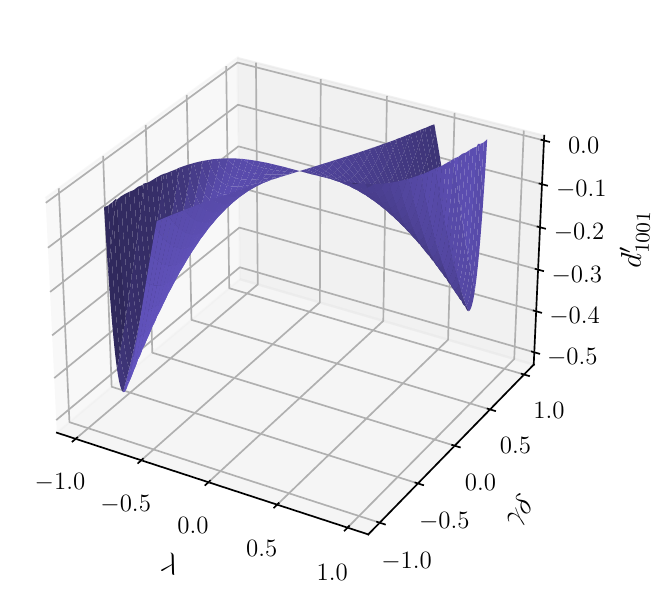}
    \caption{Witnessed region of $d'_{1001}$ for two-mode squeezed vacuum. Here, $\lambda$ is the squeezing parameter and $\gamma\delta$ is the product of coherent amplitudes of the auxiliary state, which we take to be real.}
    \label{fig:tmsvwaux}
\end{figure}

We show how the witness $d'_{1001}$ performs for the two-mode squeezed vacuum in Fig.~\ref{fig:tmsvwaux}. Here, entanglement is witnessed for all values of $\lambda\neq 0$ when paired with a suitable reference state with coherent amplitude satisfying $\gamma\delta\approx\lambda/(1-\lambda^2)$. The margin of violation also increases with increasing $|\lambda|$, reflecting the entanglement monotone behavior of the negativity.

\subsection{Two-mode Schr\"{o}dinger cat states}

\begin{figure*}
    \centering
    \includegraphics[width=1.0\linewidth]{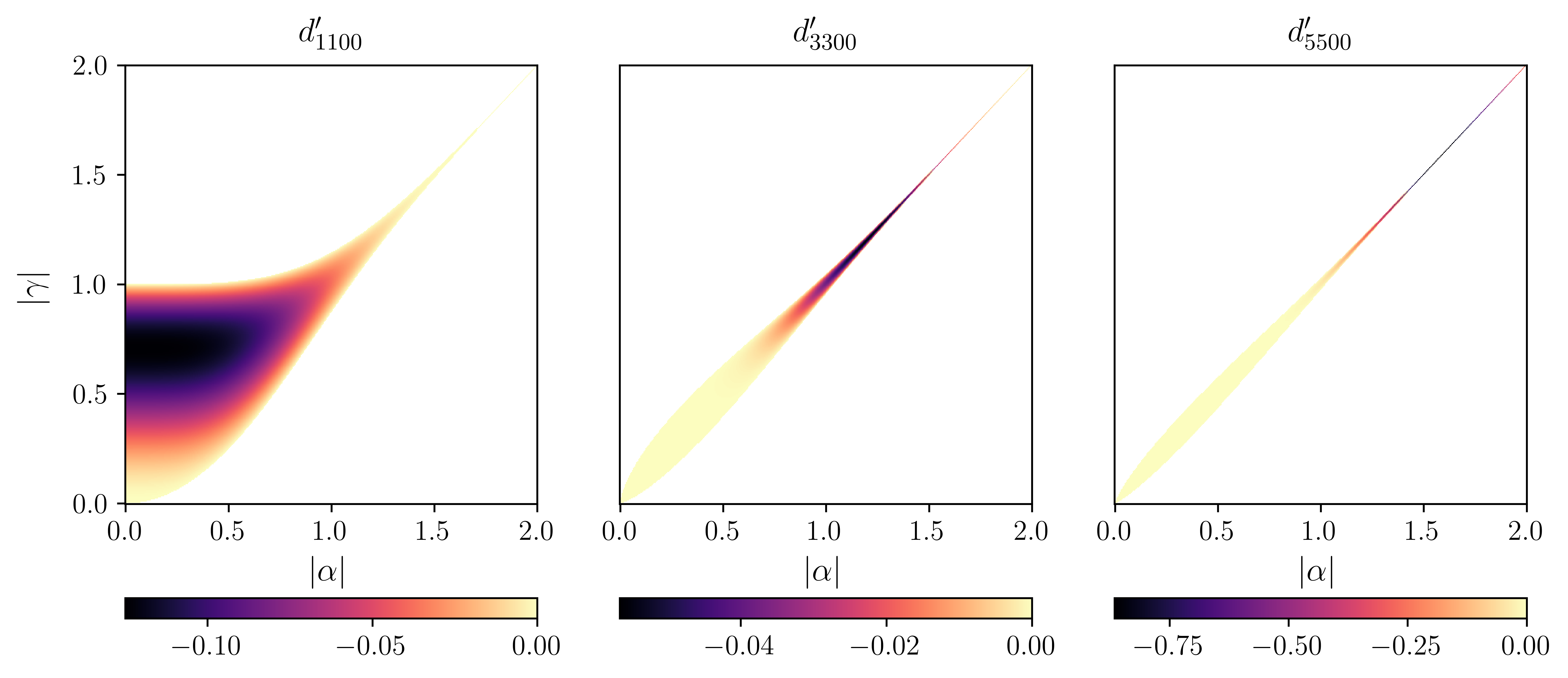}
     \caption{Witnessed regions of $d'_{mn00}$ for various values of $m$ and $n$. Here, $\alpha$ parametrizes the odd two-mode Schr\"{o}dinger cat state and $\gamma$ is the amplitude of the reference state, $\ket{\gamma,\gamma}$.}
     \label{fig:catwaux}
\end{figure*}

Another important class of bipartite-entangled states is two-mode Schr\"{o}dinger cat states. These are superpositions of coherent states and their general form is given by
\begin{equation}\label{eq:catstate}
    \ket{\Psi_\text{cat}(\alpha,\beta)}=\mathcal{N}(\alpha,\beta)\left(\ket{\alpha,\beta}+e^{\I\theta}\ket{-\alpha,-\beta}\right),
\end{equation}
where $\ket{\pm\alpha,\pm\beta}$ are the two-mode coherent states and the normalization constant is $\mathcal{N}(\alpha,\beta)=\left[2+2\cos\theta e^{-2(|\alpha|^2+|\beta|^2)}\right]^{-1/2}$. The state is entangled for all complex-valued $\alpha,\beta\neq 0$ and violates separability criteria in the form of
\begin{equation}\label{eq:cat}\begin{split}
    d'_{mnpq}&=\frac{1}{2}\left(d_{mnpq}+|\braket{\boldsymbol{a}^{\dagger m}\boldsymbol{a}^{p}\boldsymbol{b}^{\dagger q}\boldsymbol{b}^{n}}_\epsilon|^2\right)\geq0,
\end{split}\end{equation}
The terms in the minor $d_{mnpq}$ are given by the following expressions:
\begin{subequations}
\begin{equation}\begin{split}
    &\braket{\boldsymbol{a}^{\dagger m}\boldsymbol{a}^m\boldsymbol{b}^{\dagger n}\boldsymbol{b}^n}=|\alpha|^{2m}|\beta|^{2n}\begin{cases}
         1 &\text{for }m\sim n,\\
        \frac{1-\cos{\theta}e^{-\Delta}}{1+\cos{\theta}e^{-\Delta}} &\text{else},
    \end{cases}
\end{split}\end{equation}
and
\begin{equation}\begin{split}
   &\braket{\boldsymbol{a}^{\dagger m}\boldsymbol{a}^p\boldsymbol{b}^{\dagger q}\boldsymbol{b}^n}\braket{\boldsymbol{a}^{\dagger p}\boldsymbol{a}^m\boldsymbol{b}^{\dagger n}\boldsymbol{b}^q}=|\alpha|^{2m+2p}|\beta|^{2n+2q}\\
   &\qquad \qquad\times\begin{cases}
         1 &\text{for }m\sim q \textrm{ and } n\sim p,\\
        \left[\frac{1-\cos{\theta}e^{-\Delta}}{1+\cos{\theta}e^{-\Delta}}\right]^2 &\text{for }m\not\sim q \textrm{ and } n\not\sim p,\\
        \left[\frac{\sin{\theta}e^{-\Delta}}{1+\cos{\theta}e^{-\Delta}}\right]^2 &\textrm{else}.
    \end{cases}
\end{split}\end{equation}
\end{subequations}
Here, we have set $\Delta=2(|\alpha|^2+|\beta|^2)$ for ease of notation, and we define $x\sim y$ as signifying equivalence mod 2, i.e. $x=y\mod{2}$. Note again that we require either $m$ or $p$, and $n$ or $q$, to be zero. We see that the minor in Eq.~\eqref{eq:cat} can detect the presence of entanglement for $\theta\approx\pm\pi$ (where $\theta=\pi$ for odd cat states) when the conditions $m\not\sim q$ and $n\not\sim p$ are satisfied.

In Fig.~\ref{fig:catwaux} we plot $d'_{mn00}$ for an input state comprised of $\ket{\Psi_\text{cat}(\alpha,\alpha)}$ with $\theta=\pi$ and a coherent reference state $\ket{\gamma,\gamma}$. This results in the witness:
\begin{equation}\begin{split}
    d'_{mn00}&=\frac{1}{2}\bigg[|\alpha|^{2m+2n}+|\gamma|^{2m+2n}\\
    &\qquad\quad-2\text{Re}[(\alpha\gamma^*)^m(\alpha^*\gamma)^n]\coth\left(\frac{\Delta}{2}\right)\bigg]\geq 0.
\end{split}\end{equation}
The region of the parameter space that violates the criterion depends on the order of the evaluated moments: in order to witness entanglement in states with larger amplitudes $\alpha$, one has to estimate moments of increasing order. However, this becomes more difficult as $\coth{\Delta/2}$ approaches unity for large values of $\alpha$, thereby reducing the margin of violation, whilst the choice of reference state needs to be increasingly fine-tuned.

\subsection{Hermite-Gaussian wavefunction}

\begin{figure*}
    \centering
     \includegraphics[width=\textwidth]{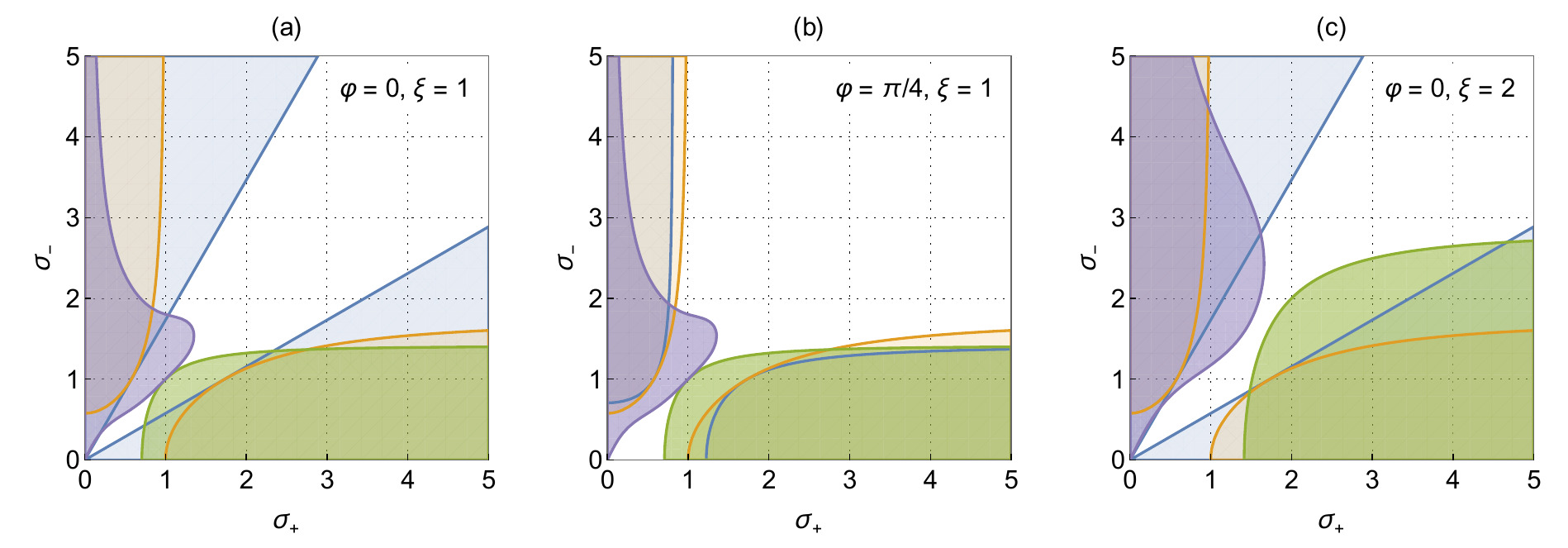}
     \caption{Comparing the regions of the parameter space witnessed by the two criteria in Eq.~\eqref{eq:hermitegaussianminors}, $d_{1001}$ (\textit{green}) and $d_{1100}$ (\textit{purple}), with the MGVT criterion \eqref{eq:mgvt} (\textit{blue}) and the second-order criterion \eqref{eq:secondorder} (\textit{orange}). We consider the original state as given in Eq.~\eqref{eq:hermitegaussian} in (a), with rotation set to $\varphi=\pi/4$ in (b) and the squeezing parameter set to $\xi=2$ in (c).}
     \label{fig:criteria}
\end{figure*}

We will now consider states described by a first-order Hermite-Gaussian wavefunction of the form
\begin{equation}\label{eq:hermitegaussian}
    \psi(x_1,x_2)=\frac{x_1+x_2}{\sqrt{\pi\sigma_-\sigma_+^3}}e^{-\frac{1}{4}\left[\left(\frac{x_1+x_2}{\sigma_+}\right)^2+\left(\frac{x_1-x_2}{\sigma_-}\right)^2\right]},
\end{equation}
where $x_1$ and $x_2$ are the canonical position coordinates and the state is entangled for all $\sigma_+,\sigma_->0$. Such states may be generated by squeezing vacuum and a single-photon Fock state by amounts $\sigma_-$ and $\sigma_+$, respectively, followed by interference on a balanced beam splitter \cite{Hertz2016}.

We restrict ourselves to minors comprised of moments that are second order in the two mode operators and that take into account correlations between the modes $\boldsymbol{a}$ and $\boldsymbol{b}$. More specifically, the state is able to violate the following separability criteria:
\begin{equation}\label{eq:hermitegaussianminors}\begin{split}
d_{1001}&=\braket{\boldsymbol{a}^\dagger\boldsymbol{a}}\braket{\boldsymbol{b}^\dagger\boldsymbol{b}}-\braket{\boldsymbol{a}^\dagger\boldsymbol{b}^\dagger}\braket{\boldsymbol{a}\boldsymbol{b}} \geq 0,\\
d_{1100}&=\braket{\boldsymbol{a}^\dagger\boldsymbol{a}\boldsymbol{b}^\dagger\boldsymbol{b}}-\braket{\boldsymbol{a}^\dagger\boldsymbol{b}}\braket{\boldsymbol{a}\boldsymbol{b}^\dagger} \geq 0.
\end{split}\end{equation}
The first inequality coincides with the criteria by Duan, Giedke, Cirac and Zoller \cite{Duan2000} if the state is centered, and hence, also for Eq.~\eqref{eq:hermitegaussian}. In the second inequality, we introduce an incremental addition in statistical analysis by introducing a fourth-order moment. The two field operators are related to the self-adjoint canonical position and momentum operators by $\boldsymbol{a}=(\boldsymbol{x}_1+\I\boldsymbol{p}_1)/\sqrt{2}$ and $\boldsymbol{b}=(\boldsymbol{x}_2+\I\boldsymbol{p}_2)/\sqrt{2}$. For further details on the analytic evaluation of the minors, see App.~\ref{app:hermitegaussian1}.

We compare the performance of these minors as entanglement witnesses to second-moment criteria: the separability criterion by Mancini, Giovannetti, Vitali and Tombesi (MGVT) \cite{Mancini2002,Giovannetti2003} and the second-moment criterion derived in Refs.~\cite{Haas2022a}. First, we introduce the locally-rotated phase space coordinates $r_i$ and $s_i$, with $i=1,2$, are defined as $r_i=x_i\cos{\varphi}+p_i\sin{\varphi}$ and $s_i=-x_i\sin{\varphi}+p_i\cos{\varphi}$ for some $\varphi\in[0,2\pi)$. Choosing to set $r_\pm=r_1\pm r_2$ and $s_\pm=s_1\pm s_2$ as the non-local position and momentum variables, we obtain the generalized MGVT criterion
\begin{equation}\label{eq:mgvt}
    \sigma_{r_\pm}^2\sigma_{s_\mp}^2\geq 1,
\end{equation}
and the second moment criterion
\begin{equation}\label{eq:secondorder}
    (\sigma_{r_\pm}^2+1)(\sigma_{s_\mp}^2+1)-\sigma^2_{r_\pm s_\mp}\geq 4,
\end{equation}
where $\sigma^2_{x}$ and $\sigma_{xy}$ are the variance of the variable $x$ and the covariance of variables $x$ and $y$, respectively, with respect to the Wigner distribution. 

In Fig.~\ref{fig:criteria}, we show how the different separability conditions compare for different values of $\sigma_+$ and $\sigma_-$. Here we have considered the use of an auxiliary state $\ket{\gamma,\delta}$ optimized over the coherent state amplitudes $\gamma$ and $\delta$. We also generalize the state in Eq.~\eqref{eq:hermitegaussian} by applying rotations and squeezing to the $(r_\pm,s_\mp)$ variables, as was done in Refs.~\cite{Haas2023b,Haas2023a}. Our minor-based witnesses and the second-order criterion \eqref{eq:secondorder} are invariant under rotations but not under squeezing, whilst the opposite applies to the MGVT criterion. We see that our witnesses are able to detect regions of the parameter space that are not flagged by the other criteria.
    
\subsection{NOON states}

Lastly, we consider the class of general pure NOON states, given by
\begin{equation}\label{eq:n00nstate}
    \ket{\Psi_\text{NOON}}=\alpha\ket{N,0}+\beta\ket{0,N},
\end{equation}
where $|\alpha|^2+|\beta|^2=1$, and the state is entangled for all integers $N\geq1$ and all complex-valued $\alpha,\beta\neq 0$. Witnessing entanglement in NOON states is highly demanding, and increasingly so for larger values of $N$. Entropic criteria valid only for pure states flag entanglement for a restricted range of excitation numbers \cite{Walborn2009,Saboia2011}, whilst all second-order and entropic criteria for mixed states fail. The Wehrl mutual information is a perfect witness for bipartite entanglement in pure states \cite{Haas2021b}, however its estimation is highly resource-intensive and does not apply for mixed states. Separability criteria based on partial transpose moments can witness entanglement for all $N$, however require three copies to access experimentally \cite{Deside2025}.

A suitable separability criterion for NOON states is given by
\begin{equation}
    d'_{00NN}=\frac{1}{2}\left(d_{00NN}+|\braket{\boldsymbol{a}^{\dagger N}\boldsymbol{b}^N}_\epsilon|^2\right)\geq 0,
\end{equation}
where
\begin{equation}\begin{split}\label{eq:d00NN}
   d_{00NN}&=\left|\begin{matrix}
        1 & \braket{\boldsymbol{a}^N\boldsymbol{b}^{\dagger N}}\\
        \braket{\boldsymbol{a}^{\dagger N}\boldsymbol{b}^N} & \braket{\boldsymbol{a}^{\dagger N}\boldsymbol{a}^N\boldsymbol{b}^{\dagger N}\boldsymbol{b}^N}\\
    \end{matrix}\right|=-|\alpha\beta|^2 (N!)^2.
\end{split}\end{equation}
In the replica case, i.e., when $\braket{\boldsymbol{a}^{\dagger N}\boldsymbol{b}^N}_\epsilon=0$, the criterion is violated for all complex-valued $\alpha,\beta\neq 0$ and for all positive integers $N$, and therefore detects entanglement perfectly. We see that the witness leverages the fact that one of the two modes is always in the vacuum state, such that $\braket{\boldsymbol{a}^{\dagger N}\boldsymbol{a}^N\boldsymbol{b}^{\dagger N}\boldsymbol{b}^N}=0$ and the minor is therefore negative. 

\begin{figure}
    \centering
    \includegraphics[width=1.0\linewidth]{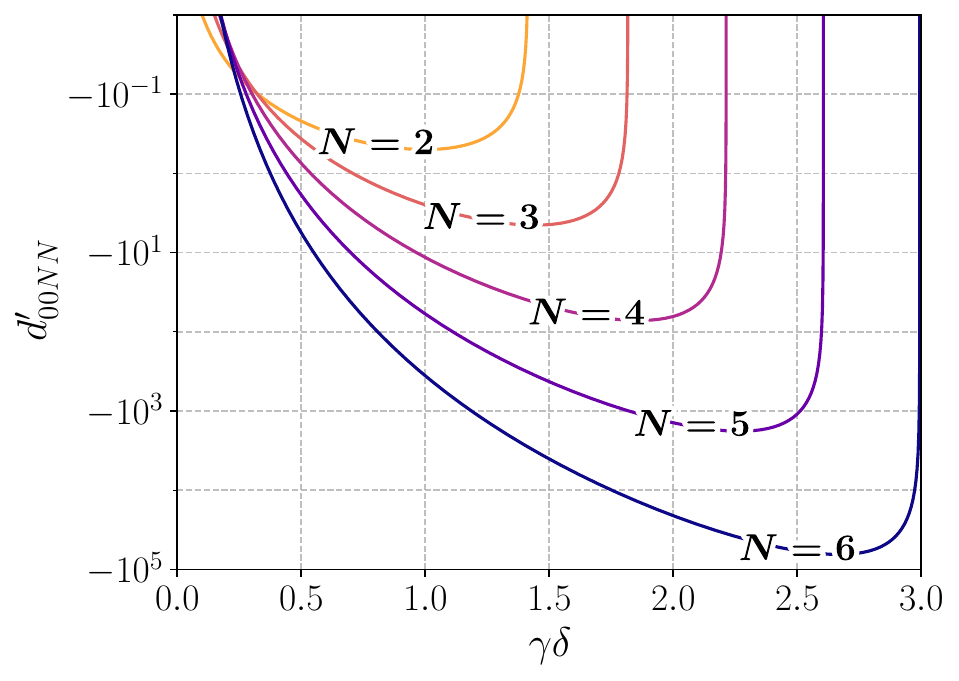}
    \caption{Minor $d'_{00NN}$ for the balanced NOON state $(\ket{N,0}+\ket{0,N})/\sqrt{2}$. Here, $N$ is the excitation number and $\gamma\delta$ is the product of coherent amplitudes of the auxiliary state, which we take to be real.}
    \label{fig:n00naux}
\end{figure}

Once again, we consider an auxiliary state of the form $\ket{\gamma,\delta}$. The criterion then transforms to
\begin{equation}\begin{split}
    d'_{00NN}&=\frac{1}{2}\left(-{|\alpha\beta|^2(N!)^2}+\left|\alpha\beta^* (N!)-(\gamma\delta^*)^N\right|^2\right)\geq 0.
\end{split}\end{equation}
In Fig.~\ref{fig:n00naux}, we show how strongly the criterion $d'_{00NN}$ is violated for different values of $N$ and coherent amplitude $\gamma\delta$. Here we emphasize that we can witness entanglement for all $N$, with the margin of violation growing larger for increasing values of $N$. Once again, this simply requires the extraction of the relevant moments without any modification to the interferometric setup. Furthermore, this is possible with the use of a separable coherent state as the auxiliary, with the scheme showing robustness in terms of the amplitudes $\gamma$ and $\delta$. This is contrast to pre-existing schemes which pose limitations in terms of, e.g., the range of valid excitation numbers $N$.

\section{Experimental considerations}\label{sec:experimentalconsiderations}

\begin{figure*}
    \centering
    \includegraphics[width=1.0\linewidth]{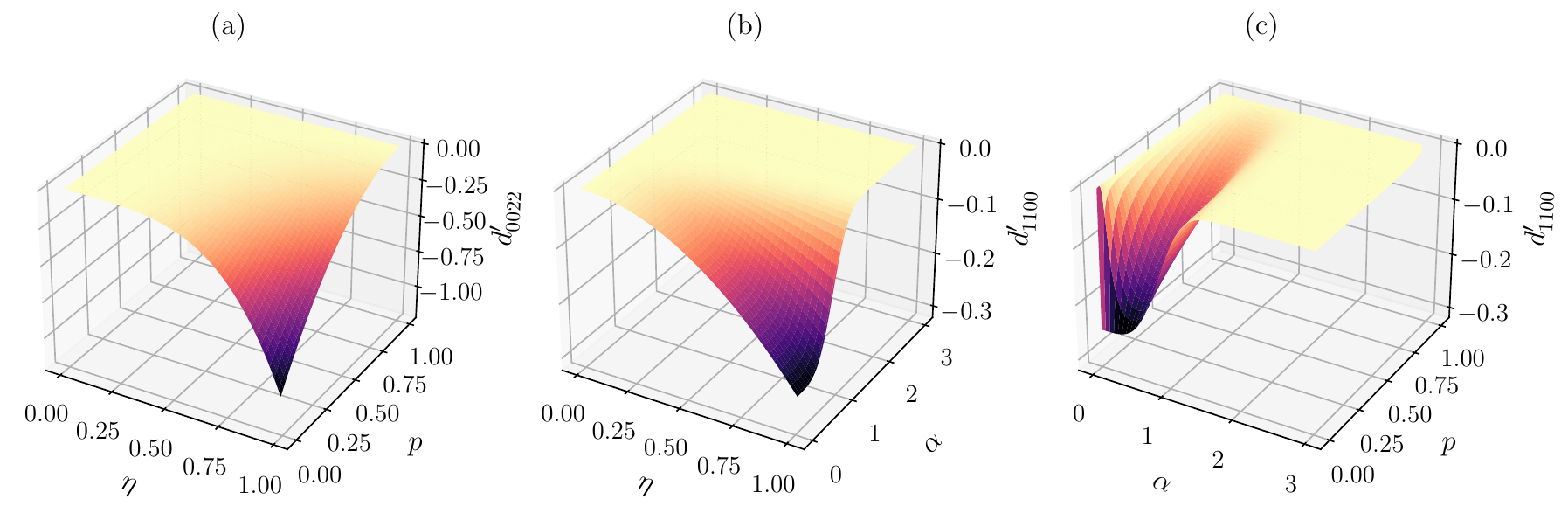}
    \caption{Witnessed regions of the separability criteria in the lossy regime for NOON states with $N=2$ (a), and two-mode Schr\"{o}dinger cat states with losses (b) and dephasing (c). Here, $\eta$ and $p$ are the loss and dephasing parameters, respectively, and $m,n$ denote the order of the moments of the minor. We set $\alpha=\beta=1/\sqrt{2}$ for the NOON state and $\alpha=\beta\in\mathbb{R}$ for the two-mode cat state. Changing the number of excitations $N$ for the former state and evaluating higher order moments for the latter state has no qualitative bearing on the performance of the measurement protocol.}
    \label{fig:losses}
\end{figure*}

\begin{figure*}
    \centering
    \includegraphics[width=0.8\linewidth]{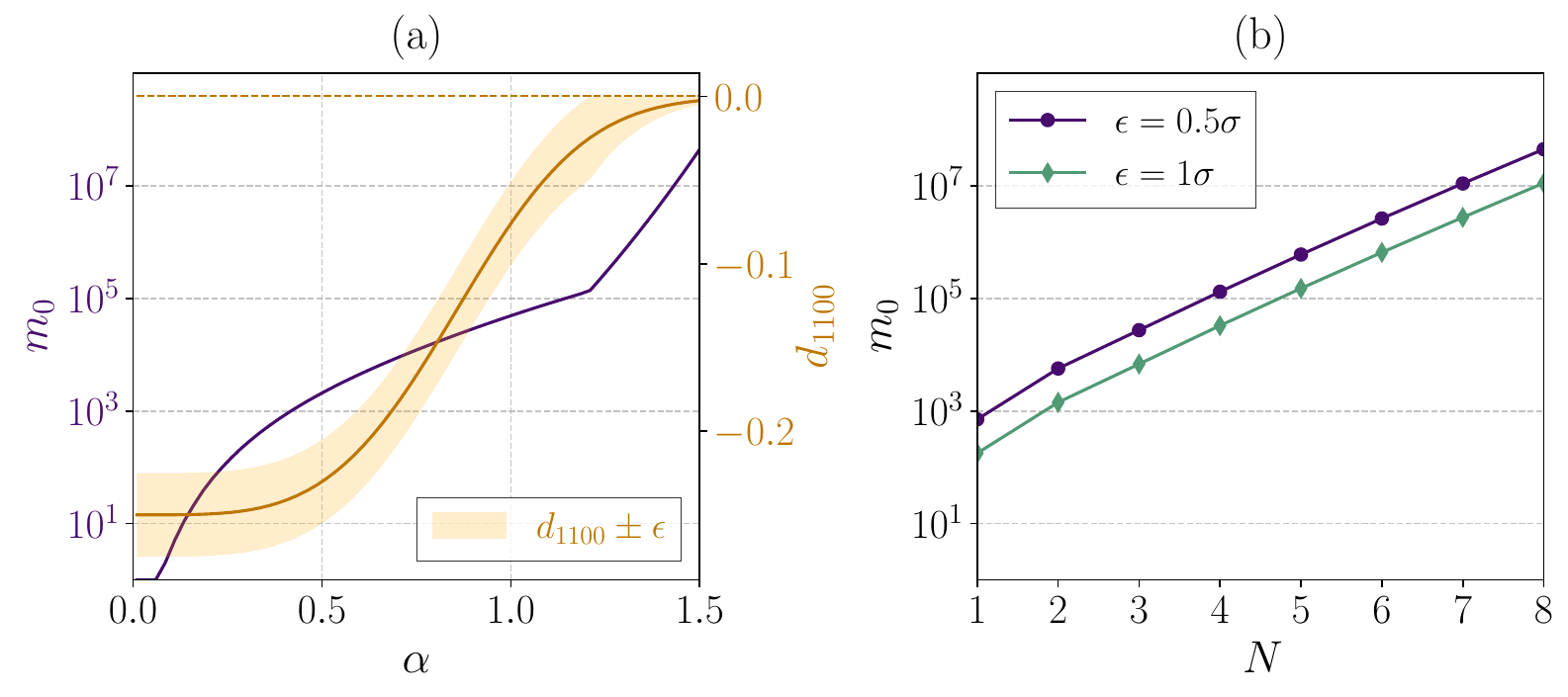}
    \caption{(a) The critical number of measurements $m_0$ needed to estimate the minor $d_{1100}$ [\textit{left axis}] and the value of the minor together with a margin of error $\epsilon=\min(0.025,|d_{1100}|)$ [\textit{right axis}] for two-mode Schr\"{o}dinger cat states $\ket{\Psi_\text{cat}(\alpha,\alpha)}$. The dotted line at $d_{1100}=0$ establishes the region below which the separability criterion is violated. Note that at $\alpha\gtrsim 1.2$ we reduce the error margin to keep the $90\%$ interval within the region of $d_{1100}<0$, leading to a kink in $m_0$. (b) The critical number of measurements $m_0$ needed to estimate the minor $d_{00NN}$ to within an accuracy level of $0.5\sigma$ and $1\sigma$ for various values of $N$. In both plots, we set the confidence $(1-\delta)$ to $90\%$.}
    \label{fig:m0}
\end{figure*}

We continue our discussion on the performance of the proposed measurement scheme to flag entanglement by accounting for experimental noise and losses, as well as the sampling complexity. In practice, the dominant mechanisms are photon losses, detector inefficiencies (where efficiencies of $\geq95\%$ have been demonstrated in superconducting nanowire detectors \cite{Reddy2019,Slussarenko2019}) and noise due to dephasing \cite{Bohmann2015,Teh2020}.

In order to describe photon losses and detector inefficiencies, we introduce beamsplitters before the photon detectors with transmissivity $\eta\in[0,1]$, such that $1-\eta$ quantifies the amount of losses that occur. The input modes of the beamsplitters are $\boldsymbol{c}$ and $\boldsymbol{r}_c$, and $\boldsymbol{d}$ and $\boldsymbol{r}_d$, and the modes transform as
\begin{equation}\label{eq:modetransformationlossy}
    \begin{pmatrix}
        \boldsymbol{c}_1\\
        \boldsymbol{d}_1\\
    \end{pmatrix} \longrightarrow\begin{pmatrix}
        \boldsymbol{c}'_1\\
        \boldsymbol{d}'_1\\
    \end{pmatrix}= \begin{pmatrix}
        \sqrt{\eta}\,\boldsymbol{c}_1 +\sqrt{1-\eta}\,\boldsymbol{r}_{c}\\
        \sqrt{\eta}\,\boldsymbol{d}_1 +\sqrt{1-\eta}\,\boldsymbol{r}_d\\
    \end{pmatrix}.
\end{equation}
We assume the absence of detector dark counts, i.e., modes $\boldsymbol{r}_c,\boldsymbol{r}_d$ have the vacuum as the input state. Under the influence of these losses, and considering the proposed measurement scheme, the evaluated minor \eqref{eq:som} becomes
\begin{equation}\begin{split}\label{eq:somlossy}
    d_{mnpq}(\eta)=&\eta_1^{m+n+p+q}\braket{\boldsymbol{a}^{\dagger m}\boldsymbol{a}^{m}\boldsymbol{b}^{\dagger n}\boldsymbol{b}^{n}}\braket{\boldsymbol{a}^{\dagger p}\boldsymbol{a}^{p}\boldsymbol{b}^{\dagger q}\boldsymbol{b}^{q}}\\
    &-\left(\frac{\eta_2}{2}\right)^{m+n+p+q}|\braket{\boldsymbol{a}^{\dagger m}\boldsymbol{a}^p\boldsymbol{b}^{\dagger q}\boldsymbol{b}^{n}}|^2 \geq 0,
\end{split}\end{equation}
where $\eta_1$ and $\eta_2$ characterize the losses occurring in either of the two measurement setups. We see that as long as $\eta_1\geq \eta_2/2$ (where both are non-negative by definition), the violation of the inequality still holds as a sufficient condition for entanglement. However, the criterion becomes weaker as losses increase.

In the case of dephasing, we introduce the dephasing parameter $p$ which takes on values in the range $[0,1]$. This determines the amount by which we reduce the coherence terms of the entangled state, effectively reducing its purity. For NOON states \eqref{eq:n00nstate}, we get
\begin{equation}\begin{split}
    \boldsymbol{\rho}_\text{NOON}&=|\alpha|^2\ket{N,0}\bra{N,0}+|\beta|^2\ket{0,N}\bra{0,N}\\
    &+(1-p)\left(\alpha\beta^*\ket{N,0}\bra{0,N}+\beta\alpha^*\ket{0,N}\bra{N,0}\right),
\end{split}\end{equation}
whilst two-mode Schr\"{o}dinger cat states \eqref{eq:catstate} become
\begin{equation}\begin{split}
    &\boldsymbol{\rho}_\text{cat}=\mathcal{N}^2(\alpha,\beta,p)\bigg[\ket{\alpha,\beta}\bra{\alpha,\beta}+\ket{-\alpha,-\beta}\bra{-\alpha,-\beta}\\
    &+(1-p)\left(e^{-\I\theta}\ket{\alpha,\beta}\bra{-\alpha,-\beta}+e^{-\I\theta}\ket{-\alpha,-\beta}\bra{\alpha,\beta}\right)\bigg].
\end{split}\end{equation}

In Fig.~\ref{fig:losses}, we show the effects of this noise on the entanglement detection scheme when using two replicas of the entangled state as the input. We assume $|\alpha|=|\beta|=1/\sqrt{2}$ in the case of NOON states, and take $|\alpha|=|\beta|$ with $\theta=\pi$ in the case of two-mode Schr\"{o}dinger cat states. We see that entanglement is still flagged for some amounts of loss and dephasing, with the margin of violation strictly monotonically decreasing in $\eta$ and $p$.

Next, we motivate the use of our proposed two-state scheme as a practically feasible method for witnessing entanglement by discussing the sampling complexity \cite{Kliesch2021}. Specifically, we consider the scaling of the number of shots required to estimate the relevant moments. The critical number of measurements $m_0$ required to achieve an estimate of some mode-operator-moment with $\epsilon$-level accuracy and confidence $(1-\delta)$ will depend on the photon-number distribution and the moments of interest.\footnote{\added{Here, $\epsilon$ is the imposed upper-bound on the difference between the estimated value and the true value, whilst $(1-\delta)$ is the confidence level with which this level of accuracy is achieved.}} Whilst entanglement in Gaussian and Hermite--Gaussian states can be detected using low moments, the estimation of which is sample-efficient, high moments may be required for certain parameters of strongly non-Gaussian states. This includes the family of entangled cat states as well as NOON states, which we consider in this work. For further details, refer to App.~\ref{app:samplingcomplexity}.

The two-mode Schr\"{o}dinger cat state has an unbounded photon-number distribution, and hence we make use of Chebyshev's inequality in order to establish the critical number of measurements, $m_0$. This is given by
\begin{equation}
    m_0=\frac{\text{Var}(\boldsymbol{O})}{\delta\,\epsilon^2},
\end{equation}
where $\boldsymbol{O}$ is the operator for which we wish to estimate the expectation value. The measurement of the minor $d_{mnpq}$ for such cat state, given in Eq.~\eqref{eq:cat}, reduces to the estimation of a sum of expectation values when $p,q=0$. Given that the measurement shots are carried out on uncorrelated copies of the state, we may add the variances of the individual operators to establish the variance for the estimate of the minor. 

In the case of NOON states, we seek to estimate the minor $d_{00NN}$, given in Eq.~\eqref{eq:d00NN}. Here, the number distribution is bounded from above: the only measurement outcome of $\boldsymbol{a}^{\dagger N}\boldsymbol{a}^{ N}\boldsymbol{b}^{\dagger N}\boldsymbol{b}^{ N}$ (and all operators in its photon-number operator decomposition) acting on the NOON state is zero, whilst $(\boldsymbol{c}^\dagger\boldsymbol{c}\,\boldsymbol{d}^\dagger\boldsymbol{d})^N$, which is used to evaluate the second summand of the minor, is bounded from above by $N^{2N}$. We may therefore apply Hoeffding's inequality, from which we obtain
\begin{equation}
    m_0=\frac{(2N+1)^2N^{4N}}{2\epsilon^2}\ln{\frac{2}{\delta}}.
\end{equation}
Note that the factor of $(2N+1)^2$ here is due to the minimum number of evenly-spaced points from the parameter space of $\phi,\phi'$ that need to be sampled in order to prevent aliasing \cite{Shannon1949}.

In Fig.~\ref{fig:m0}, we show how $m_0$ scales with $\alpha$ in the case of two-mode Schr\"{o}dinger cat state, $\ket{\Psi_\text{cat}(\alpha,\alpha)}$, as well for NOON states with relatively large excitation numbers, together with their respective margin of errors. In all cases, we consider a confidence level of $90\%$. Although, as may be expected, we see near-exponential scaling, we still can establish violation of the PPT criterion to a high level of accuracy for an extensive range of parameters under realistic shot numbers. Using up to $m_0\approx 10^6$ samples, one can witness entanglement in cat states for $\alpha\lesssim 1.25$, and in NOON states with $N\leq5$.

\addedd{Finally, we address the limitation due to the current unavailability of ideal PNR detectors. Here, finite-resolution and multiplexed PNR detectors are sufficient for experimental implementation. The entanglement witness becomes weaker as the order of the estimated minors increases and the photon number distribution broadens. However, for states with a sufficiently low photon number, the strength of the witness does not decrease significantly. Note that in order to preserve the strength of the witness, the largest photon number that can be reliably detected needs to be equal to the largest photon number contribution to the state arriving at the detector.}

\section{Conclusions}\label{sec:conclusions}

We have presented a measurement scheme requiring only two interfering states and photon-number-resolving detection in order to detect bipartite entanglement in continuous-variable systems. The photon-number correlations that are measured at the output of the interferometer contain information about moments of the states' field operators, which in turn can be extracted by simple Fourier analysis. Hence, we can use the scheme to show violation of certain separability criteria derived from the Shchukin--Vogel hierarchy, which consists of separability conditions comprised of such moments.

Moreover, we have shown that the states do not have to be two copies of the entangled state; instead, one can employ an easier-to-source auxiliary state, such as a product coherent state. To do so, we construct a modified criterion in terms of the observables that can be evaluated with the measurement protocol and show that it is a necessary separability condition. The scheme is shown to be able to flag entanglement, even when using a coherent state as a reference, of two-mode squeezed vacuum, states with a Hermite-Gaussian wavefunction, two-mode Schr\"{o}dinger cat states, and NOON states with arbitrary excitation number $N$. Finally, we also addressed some possible limitations in the form of experimental noise and sampling complexity. In terms of experimental feasibility, states which require the estimation of higher-order moments generally become harder to detect. Nonetheless, we have shown that entanglement detection in, for example, two-mode Schr\"{o}dinger cat states and NOON states is possible for a wide range of state parameters. Optical losses, as expected, reduce the violation of the entanglement criteria. However, in the regime of moderate amounts of loss, the scheme is still robust and does not flag entanglement falsely.

In this work, we have focused on one class of components \eqref{eq:set} making up the photon-number correlation functions, as the mapping of these elements to the Shchukin--Vogel hierarchy is quite straightforward. However, it would be interesting to see whether photon-number measurements could reveal the presence of entanglement of some broader class of states by considering separability criteria formulated in terms of other components. On a similar note, we have tried to extend the protocol to capture higher-order minors. This has proved to be not straightforward, as we rely on the use of \emph{two} uncorrelated input states to factorize expectation values into a product of \emph{two} terms. Future work could explore alternative ways of extending this work to a broader class of moments and with potentially more interfering states.

\section*{Acknowledgments}
T.H. acknowledges support from the F.R.S. -- FNRS under project CHEQS within the Excellence of Science (EOS) program.
This research is supported by funding from the German Research Foundation (DFG) under the project identifier 398816777-SFB 1375 (NOA).

\appendix

\section{Hermite--Gaussian wavefunction}\label{app:hermitegaussian1}

Analytic expressions for the minors are obtained by consideration of $\boldsymbol{a}=(\boldsymbol{x}_1+\I\boldsymbol{p}_1)/\sqrt{2}$ and $\boldsymbol{b}=(\boldsymbol{x}_2+\I\boldsymbol{p}_2)/\sqrt{2}$, and using the definition
\begin{equation}
    \braket{\boldsymbol{f}(\vec{\boldsymbol{x}},\vec{\boldsymbol{p}})}=\int\mathrm{d}\vec{x}\,\mathrm{d}\vec{p}\,W(\vec{x},\vec{p})\,\tilde{f}(\vec{x},\vec{p}),
\end{equation}
where $W(\vec{x},\vec{p})$ is the Wigner distribution of the state and $\tilde{f}(\vec{x},\vec{p})$ is the Wigner--Weyl transform of the operator $\boldsymbol{f}(\vec{\boldsymbol{x}},\vec{\boldsymbol{p}})$. For the wavefunction in Eq.~\eqref{eq:hermitegaussian}, we obtain the following non-vanishing moments:
\begin{subequations}
    \begin{equation}
        \braket{\boldsymbol{x}_1^2}=\braket{\boldsymbol{x}_2^2}=\frac{-\sigma_-^2+3\sigma_+^2}{\sigma_-^2+3\sigma_+^2}\braket{\boldsymbol{x}_1\boldsymbol{x}_2}=\frac{\sigma_-^2+3\sigma_+^2}{4}
    \end{equation}
    \begin{equation}
        \braket{\boldsymbol{p}_1^2}=\braket{\boldsymbol{p}_2^2}=\frac{-\left(\sigma_+^2+\sigma_-^2\right)}{\sigma_+^2+3\sigma_-^2}\braket{\boldsymbol{p}_1\boldsymbol{p}_2}=\frac{\sigma_+^2+3\sigma_-^2}{4\sigma_+^2\sigma_-^2}
    \end{equation}
    \begin{equation}
        \braket{\boldsymbol{x}_1^2\boldsymbol{x}_2^2}=\frac{3}{16}\left(5\sigma_+^4 -2\sigma_-^2\sigma_+^2+\sigma_-^4 \right)
    \end{equation}
    \begin{equation}
        \braket{\boldsymbol{x}_1^2\boldsymbol{p}_2^2}=\braket{\boldsymbol{p}_1^2\boldsymbol{x}_2^2}=\frac{3}{16}\frac{(\sigma_+^2+\sigma_-^2)^2}{\sigma_+^2\sigma_-^2}
    \end{equation}
    \begin{equation}
        \braket{\boldsymbol{p}_1^2\boldsymbol{p}_2^2}=\frac{3}{16}\frac{(3\sigma_+^4+10\sigma_+^2\sigma_-^2-\sigma_-^4)}{\sigma_-^4\sigma_+^4}
    \end{equation}
\end{subequations}

\section{Sampling complexity}\label{app:samplingcomplexity}

For a detailed review on quantum state certification, we refer the reader to Ref.~\cite{Kliesch2021}. Let the empirical mean estimator $Y^{(m)}$ of the moment $\braket{\boldsymbol{O}}$ be
\begin{equation}
    Y^{(m)}=\frac{1}{m}\sum_{i=1}^m O^{(i)},
\end{equation}
where $O^{(i)}$ is the $i$th measurement outcome of operator $\boldsymbol{O}$. Here we assume sampling from independent and identically-distributed states.

In general, we may consider Chebyshev's inequality:
\begin{equation}\label{eq:chebyshev}\begin{split}
    &\Pr\left[\left| Y^{(m)}-\mathbb{E}[ Y^{(m)}]\right|\geq k \mathbb{E}\left[\left|Y^{(m)}-\mathbb{E}\left[ Y^{(m)}\right]\right|^n\right]^{1/n}\right]\\
    &\qquad\qquad\leq\frac{1}{k^n},
\end{split}\end{equation}
where $k>0$ and $n\geq 2$. Setting $n=2$, $\delta$ to be an upper-bound of the RHS (i.e. $1/k^2 \leq \delta$) and $\epsilon=k \mathbb{E}\left[\left|Y^{(m)}-\mathbb{E}\left[ Y^{(m)}\right]\right|^n\right]^{1/n}$, one obtains
\begin{equation}\begin{split}
    m \geq \frac{\text{Var}(\boldsymbol{O})}{\epsilon^2 \delta}.
\end{split}\end{equation}
Hence, we obtain a lower bound on the number of measurements $m$ required to ensure the specified level of accuracy according to Eq.~\eqref{eq:chebyshev}, with the inequality saturated at the critical number of measurements $m_0$.

For bounded eigenspectra, $\text{spec}(\boldsymbol{O})\subseteq[a,b]$, one may apply Hoeffding's inequality for improved bounds:
\begin{equation}
    \Pr[|Y^{m} - \mathbb{E}[Y^{m}]|\geq \epsilon]\leq 2\exp\left(-\frac{2\epsilon^2m}{(b-a)^2}\right).
\end{equation}
Again, setting $\delta$ to be an upper-bound for the inequality, one obtains an improved lower bound for the number of measurements $m$ required for a given level of accuracy:
\begin{equation}
    m\geq m_0=\frac{(b-a)^2}{2\epsilon^2}\ln{\frac{2}{\delta}}.
\end{equation}

\bibliography{bibliography}

\end{document}